# Tuning the Threshold Voltage of MoS$_2$ Field-Effect Transistors *via* Surface Treatment

Wei Sun Leong,[#,†] Yida Li,[#,†] Xin Luo,[#,‡,¥] Chang Tai Nai,[§] Su Ying Quek,[*,‡,¥] and John T. L. Thong[*,†]

[†]Department of Electrical and Computer Engineering, National University of Singapore, Singapore 117583.

[‡]Department of Physics, Centre for Advanced 2D Materials and Graphene Research Centre, National University of Singapore, Singapore 117546.

[¥]Institute of High Performance Computing, 1 Fusionopolis Way, #16-16 Connexis, Singapore 138632.

[§]Department of Chemistry, National University of Singapore, Singapore 117543.

*Address correspondence to elettl@nus.edu.sg (experiment); phyqsy@nus.edu.sg (theory)

[#]These authors contributed equally to this work.

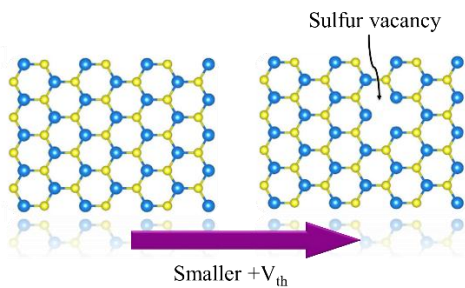

Sulfur vacancy

Smaller +V$_{th}$

**ABSTRACT**

Controlling the threshold voltage ($V_{th}$) of a field-effect transistor is important for realizing robust logic circuits. Here, we report a facile approach to achieve bidirectional $V_{th}$ tuning of molybdenum disulfide ($MoS_2$) field-effect transistors. By increasing and decreasing the amount of sulfur vacancies in the $MoS_2$ surface, the $V_{th}$ of $MoS_2$ transistors can be left- and right-shifted, respectively. Transistors fabricated on perfect $MoS_2$ flakes are found to exhibit two-fold enhancement in mobility and a very positive $V_{th}$ (18.5 ± 7.5 V). More importantly, our elegant hydrogen treatment is able to tune the large $V_{th}$ to a small value (~0 V) without any performance degradation simply by reducing the atomic ratio of S:Mo slightly; in other words, creating a certain amount of sulfur vacancies in the $MoS_2$ surface, which generate defect states in the band gap of $MoS_2$ that mediate conduction of a $MoS_2$ transistor in the subthreshold regime. First-principles calculations further indicate that the defect band's edge and width can be tuned according to the vacancy density. This work not only demonstrates for the first time the ease in tuning the $V_{th}$ of $MoS_2$ transistors, but also offers a process technology solution that is critical for further development of $MoS_2$ as a mainstream electronic material.

**Introduction**

Among two-dimensional layered materials, transition metal dichalcogenides (TMDs) are being considered for their potential as active device materials due to their exceptional electronic and optical properties. Molybdenum disulfide ($MoS_2$), in particular, has been extensively studied for a number of years. The use of $MoS_2$ as the channel material in field-effect transistors (FETs) has been theoretically predicted to bestow several attractive traits that include reasonable room temperature mobility, almost ideal switching characteristics, and superior on/off current ratio ($>10^{10}$) due to the large direct bandgap (1.9 eV) for monolayer $MoS_2$.[1-3] Nevertheless, despite all the hype about $MoS_2$ being a promising candidate to replace silicon in such devices, performance inadequacies such as large variations in threshold voltage ($V_{th}$) for $MoS_2$ transistors represent the present-day reality. As a reference, the $V_{th}$ of silicon-based semiconductor technology at the 22 nm technology node is 0.289 V for high performance logic and 0.413 V for low operating power logic.[4] In contrast, values of $V_{th}$ of back-gated $MoS_2$ FETs being reported vary from -30 V to 40 V.[5-16] Considering the ultimate goal of implementing $MoS_2$ in future electronic devices, the ability to adjust the $V_{th}$ of $MoS_2$ FETs to the desired level is of paramount importance.

Compared to conventional silicon-based transistors based on 3D bulk semiconductor channels, the physical origin of the $V_{th}$ for 2D FETs is quite different. In the 3D case, an inversion layer forms under the gate to form a conducting channel as a result of long-range band bending within the bulk material from the substrate towards the gate. In the 2D case, the concept of band bending in the direction normal to the gate electrode does not apply; instead, a conducting channel is formed when the energies of states in $MoS_2$ that contain majority carriers (*i.e.* conduction band) become aligned with the Fermi level of the source/drain electrodes.[17, 18] This alignment is achieved by the gate voltage which induce shifts of the energy levels in the $MoS_2$ channel relative to the Fermi level in the source/drain electrode, which is fixed by a constant drain-to-source voltage.

Here, we demonstrate that the $V_{th}$ of $MoS_2$ FETs can be easily tuned bi-directionally by controlling the amount of sulfur vacancies on the $MoS_2$ surface. The sulfur vacancies (or defects) in the basal plane of MoS$_2$ create states (*i.e.* defect bands) within the band gap that have been shown to mediate conduction in the subthreshold regime,[19, 20] and these defect levels can be tuned according to the vacancy density. In this paper, we show that sulfur and hydrogen treatments are effective in right- and left- shifting of the V$_{th}$ of MoS$_2$ transistors, respectively. Specifically, the former treatment reduces sulfur vacancies that exist even in the intrinsic MoS$_2$ flakes exfoliated from natural crystal and which are not uniformly distributed.[20, 21] By annealing the freshly-exfoliated MoS$_2$ flakes in a sulfur-rich environment, the dangling bonds of Mo are expected to bond readily with the sulfur atoms to reduce the density of sulfur vacancies in the MoS$_2$ surface. On the other hand, the latter treatment creates sulfur vacancies in the basal plane of MoS$_2$ through a gasification process: S (solid) + H$_2$ (gas) → H$_2$S (gas)[22] and this chemical reaction is unlikely to result in the formation of double vacancies in the MoS$_2$ from both thermodynamic and kinetic points of view as predicted theoretically.[23] In other words, this chemical reaction does not require defects (*i.e.*, sulfur vacancies) in the basal plane of MoS$_2$ as initiation sites. Thus far, for most of the reported FETs fabricated on pristine MoS$_2$ flakes, the V$_{th}$ falls into the negative regime, while our FETs fabricated using sulfur-treated MoS$_2$ flakes exhibit very positive V$_{th}$ with respectable performance including high on-off current ratio (> $10^5$) and 2-fold enhancement in electron mobility compared to that of intrinsic MoS$_2$ flakes. More importantly, we also demonstrate that the large positive V$_{th}$ of these MoS$_2$ FETs can easily be tuned to a smaller value (in the opposite direction) without any degradation in terms of electron mobility and on/off current ratio. This work not only sheds light on the ease in tuning the V$_{th}$ of MoS$_2$ FETs, but also offers a promising approach that is compatible with silicon CMOS fabrication technology for threshold voltage tuning in the development of MoS$_2$ devices.

**Results and Discussion**

In order to demonstrate that the V$_{th}$ of MoS$_2$ transistors can be tuned by modifying the amount of sulfur vacancies, we fabricated more than 50 back-gated FETs on 2 types of exfoliated MoS$_2$ flakes: pristine (as-exfoliated) and surface-treated. For this study, all MoS$_2$ flakes were first exfoliated from molybdenite crystal (SPI supplies®) on oxidized degenerately p-doped silicon substrates with 285 nm thick SiO$_2$. It should be noted that all MoS$_2$ flakes selected for the FETs fabrication are of similar thickness (5 nm) and uniform width (2-3 μm) as confirmed by atomic force microscopy (Figure S1). After that, one-third of the selected MoS$_2$ flakes were preserved as control samples, while the remaining parts were subjected to the surface treatment of using either sulfur (Figure S2) or hydrogen (see Experimental for details). Subsequently, transistors were fabricated on each MoS$_2$ flake (Figure 1a) with the same device dimensions, where the channel length and contact length are both 1 μm, and the channel width and contact width are about 2-3 μm, being the natural width of the exfoliated MoS$_2$ flakes. Electrical measurements for all MoS$_2$ transistors were conducted in a high vacuum chamber at room temperature. Throughout this work, we have adopted the most common $V_{th}$ extraction method for semiconductor device analysis, which is the extrapolation in the linear region (ELR) method.[24] In the ELR method, the $V_{th}$ is defined by the gate voltage axis intercept of the linear extrapolation of the transistor's transfer characteristic ($I_D - V_G$) curve at its maximum first derivative point, which is also the point of maximum transconductance, $g_m = dI_D/dV_G$.

We observe significant difference in the $V_{th}$ of MoS$_2$ transistors fabricated on both pristine and surface-treated MoS$_2$ flakes. Figure 1b shows the average $V_{th}$ of transistors fabricated on different types of MoS$_2$ flakes and their respective on/off current ratio. As can be seen in Figure 1b, the $V_{th}$ of the transistors fabricated on pristine MoS$_2$ flakes falls in the negative regime (-15 ± 5 V) with an average on/off current ratio of 10$^5$. On the other hand, the $V_{th}$ of transistors fabricated on hydrogen-treated MoS$_2$ flakes, that are foreseen to contain more sulfur vacancies compared to that of pristine MoS$_2$ flakes, is located in a more negative regime (-30.3 ± 5.7 V) but the average on/off current ratio of this group of transistors is about an order of magnitude smaller than that of pristine MoS$_2$ flakes, which could be attributed to the chemical reaction between the electropositive Ti electrodes and the hydrogen-treated MoS$_2$ surface that contains many sulfur vacancies prior to formation of Ti contacts.[25] In contrast, the $V_{th}$ of transistors fabricated on sulfur-treated MoS$_2$ flakes that are foreseen to have much fewer sulfur vacancies compared to that of pristine MoS$_2$ flakes shows not only a right-shift, but a very positive value (18.5 ± 7.5 V). Furthermore, it is worth noting that the average on/off current ratio of transistors fabricated on sulfur-treated MoS$_2$ flakes is comparable to that of transistors fabricated on pristine MoS$_2$ flakes (Figure 1b).

Despite having a right-shift in the $V_{th}$ without on/off ratio degradation, our MoS$_2$ transistors fabricated on sulfur-treated MoS$_2$ flakes exhibit a two-fold improvement in mobility (in the linear regime) compared to those fabricated on pristine MoS$_2$ flakes. For all MoS$_2$ transistors fabricated on both pristine and surface-treated MoS$_2$ flakes, we extracted and compared the maximum field-effect mobility *via* equation (1):

$$\mu = \frac{1}{C_{ox} \cdot V_{DS}} \cdot \frac{L_{ch}}{W_{ch}} \cdot \frac{\Delta I_{DS}}{\Delta V_g} \quad (1)$$

where $C_{ox}$ is the gate capacitance ($1.21 \times 10^{-8}$ F/cm$^2$ for 285 nm thick SiO$_2$),[26] $L_{ch}$ and $W_{ch}$ are channel length and width, respectively, $I_{DS}$ is the drain current, $V_{DS}$ is the drain voltage and $V_g$ is the gate voltage.

It should be noted that all of our MoS$_2$ transistors exhibit n-type behaviour (conduction mainly in the positive $V_g$ regime) regardless of whether the exfoliated MoS$_2$ flakes are pristine, sulfur-treated or hydrogen-treated. The average value of the extracted electron mobility for the transistors fabricated on pristine MoS$_2$ flakes is $37.6 \pm 8.4$ cm$^2$/V-s, and $27.5 \pm 7.2$ cm$^2$/V-s for transistors fabricated on hydrogen-treated MoS$_2$ flakes. On the other hand, the average mobility of the transistors fabricated on the sulfur-treated MoS$_2$ flakes is $85.9 \pm 12.6$ cm$^2$/V-s, which is about twice better than that of the pristine MoS$_2$ flakes with similar thickness (5 nm). We attribute this two-fold improvement in electron mobility (in the linear regime) to the reduction of defects (*i.e.* sulfur vacancies) in the basal plane of MoS$_2$ arising from the sulfur treatment. This is consistent with band-like transport in the linear regime, where defects act as scattering centers.[19, 27] In addition, the $V_{th}$ for all MoS$_2$ transistors presented in Figure 1b remains the same even after 2 weeks of storage in ambient conditions (25 °C, 1 atm), regardless of whether the exfoliated MoS$_2$ flakes are pristine or surface-treated (Figure S3).

We then explored the potential of further tuning the $V_{th}$ of the finished MoS$_2$ transistors by using a hydrogen treatment which is expected to increase the amount of sulfur vacancies in the basal plane of the MoS$_2$ channel. For this study, 10 FETs fabricated on sulfur-treated MoS$_2$ flakes were selected and subjected to a hydrogen treatment. Thereafter, all of the 10 MoS$_2$ transistors were again electrically tested in a high vacuum environment at room temperature. The extracted $V_{th}$ and on/off current ratio for this group of MoS$_2$ transistors are plotted in Figure 1b for clear comparison. Interestingly, as can be seen in Figure 1b, the set of transistors fabricated on sulfur-treated MoS$_2$ flakes initially exhibits a clear left-shifting of V$_{th}$ from 18.5 ± 7.5 V to -2 ± 5 V following the hydrogen post-treatment, while maintaining the on/off current ratio for the MoS$_2$ transistors as before. In short, the results imply that the V$_{th}$ of MoS$_2$ transistors can be tuned by controlling the amount of sulfur vacancies on the surface of the uncovered MoS$_2$ channel and the proposed hydrogen treatment at room temperature is eminently suitable as a post-processing treatment for fine tuning the V$_{th}$ of MoS$_2$ transistors.

In Figure 2, we compare the transfer characteristics of a MoS$_2$ transistor that was first fabricated on a sulfur-treated MoS$_2$ flake and electrically tested once followed by a hydrogen post-treatment and electrical testing (the output characteristics are shown in Figure S4a). The typical MoS$_2$ transistor exhibits n-type behaviour with a threshold voltage value of 12 V when it was freshly-fabricated, and 0 V after the hydrogen post-treatment. This represents a clear left-shifting of the V$_{th}$ for the finished MoS$_2$ transistor to a smaller value by the creation of a large number of sulfur vacancies in the basal plane of the uncovered MoS$_2$ channel. Furthermore, the field-effect mobility for the MoS$_2$ transistor extracted at maximum transconductance in the linear regime *via* equation (1) is 95.3 cm$^2$/V-s as-fabricated and remains comparable even after the hydrogen post-treatment (98.2 cm$^2$/V-s). This suggests that the density of sulfur vacancies generated using such treatment, while sufficient to tune V$_{th}$ to a very small value, is still small enough not to cause any significant degradation in mobility in the linear regime. In addition, the on/off current ratio (~10$^5$) for the MoS$_2$ transistor after the hydrogen post-treatment does not vary much as compared to that of the as-fabricated transistor (Figure 2b). This observation again reinforces the suitability of our room temperature treatment in preserving the electrical performance of MoS$_2$ transistors. In brief, the results signify that the V$_{th}$ for a finished MoS$_2$ transistor can be tuned to a smaller value by increasing the amount of sulfur vacancies in the uncovered MoS$_2$ channel without any degradation in terms of electron mobility and on/off current ratio.

To understand the mechanism of how the V$_{th}$ of a finished MoS$_2$ transistor can be tuned to a smaller value by increasing the amount of sulfur vacancies in the MoS$_2$ channel, we perform density functional theory (DFT) calculations to model the essential physics in these transistors. In these calculations, we consider only bilayer MoS$_2$ films as the band gap and qualitative band structure features do not change significantly when the number of MoS$_2$ layers increases from two layers[28] and most of the current flows through the top few layers in a MoS$_2$ transistor.[29] As discussed earlier, an applied gate voltage will shift the energy levels in the MoS$_2$ channel regions, while the voltages of the source/drain electrodes are fixed according to the experiment. Here, we consider the Ti-covered-MoS$_2$ system as the source/drain electrode in a transistor configuration, because Ti binds strongly to MoS$_2$, with metal-induced gap states in the MoS$_2$ band gap (Figure S5a); the metallic screening in the Ti-covered-MoS$_2$ system implies that a gate voltage will not affect the energy levels of this system. Due to charge redistribution,[16, 30-32] the work function of Ti-covered-MoS$_2$ is 5.1 eV, in contrast to 4.3 eV computed for pure Ti. The Ti-covered-MoS$_2$ system is an appropriate model electrode for the S and S+H cases (Figure 1b) as the sulfur treatment removes most of the vacancies in MoS$_2$, while the room temperature hydrogen treatment should not affect the Ti or Ti-covered-MoS$_2$ region. We first consider transistors that are in the off-state when the back gate voltage $V_g = 0$. In that case, a large onset of current will occur when the applied gate voltage is sufficient to align the defect levels in the MoS$_2$ channel to the Fermi level in the Ti-covered-MoS$_2$ electrode, as illustrated by Figures 3a and 3b. We call this energy shift (energy difference between the defect band edge and the Fermi level) as ΔE. We compute ΔE by determining the Fermi level position in Ti-covered-MoS$_2$ electrode, and comparing this to the defect bands' edge in the MoS$_2$ channel (Table S1). Different densities of sulfur vacancies ($N_{Vs}$) are modeled by removing S atoms in different sized MoS$_2$ supercells, for example, a defect density of 4.7 x 10$^{13}$ cm$^{-2}$ can be achieved by removing one S atom in a 5×5 MoS$_2$ supercell. It should be noted that the range of $N_{Vs}$ (~10$^{13}$ cm$^{-2}$) modelled in this work was chosen based on some earlier findings about sulfur vacancies on the MoS$_2$ flakes exfoliated from natural MoS$_2$ bulk crystal through transmission electron microscopy (TEM) statistical analysis[20, 27, 33] and high-resolution scanning tunneling microscopy (STM) studies.[21, 34]

The energy band diagrams in Figure 3 are obtained by extracting the DFT calculated energy levels of both Ti-covered-MoS$_2$ electrode and MoS$_2$ channel regions (Figure S5). For the vacancy density considered in Figures 3a and 3b ($N_{V_S}$ = 4.7 x $10^{13}$ cm$^{-2}$), the transistor is in the off-state at $V_g$ = 0, and the defect levels in MoS$_2$ channel must be shifted down to align with the Fermi level in the source/drain electrodes, *i.e.* $V_{th}$ > 0. Comparing cases 1 and 2 in Figure 3c, we see that the defect band edge shifts down with increasing density of sulfur vacancies $N_{V_S}$, implying that as long as the transistor is still in the off-state at $V_g$ = 0 (*i.e.* the Fermi level in the source/drain electrodes is below the defect level), ΔE is smaller for larger vacancy densities, which means smaller positive $V_{th}$ is required, in good agreement with experiment (Figure 1b).

Although ΔE is predicted to be only 0.1 and 0.2 eV for cases 1 and 2 respectively, the change in ΔE with vacancy density is robust. This trend is consistent across different vacancy densities as summarized in Table S1. Furthermore, the values of ΔE as well as the corresponding differences are 10 times larger than the precision in our calculations; in particular, we have also verified that ΔE and other relevant energy levels do not change with higher kinetic energy cutoffs and denser Brillouin Zone samplings, indicating that these values are converged in our calculations (Table S2 and Figure S6). In addition, the correspondence between experimentally applied gate voltages and theoretically computed values of ΔE is consistent with a recent gated scanning tunneling spectroscopy study, where Lu *et al.* demonstrated that the position of the Fermi energy relative to the conduction band edge in the MoS$_2$ channel of a back-gated transistor does not vary more than 0.1 eV for $V_g$ ranges from -15 V to 40 V;[18] in particular, the thickness of their MoS$_2$ flake and transistor configuration are similar to that of this work.

Our measurements on the group of S+H transistors give a $V_{th}$ of -2 ± 5 V (as indicated in Figure 1b), meaning that the $V_{th}$ is very small, and can sometimes be positive or negative depending on which transistor is being measured. The small positive $V_{th}$ corresponds to a case where ΔE as discussed above is extremely small, consistent with a vacancy density $N_{V_S}$ similar to or slightly larger than 13 x $10^{13}$ cm$^{-2}$ (case 1). When $V_{th}$ is small and negative, the transistor is already in the on-state when $V_g$ = 0 (*e.g.* see the red curve in Figure 2a for S+H), indicating that the Fermi level in the source/drain electrodes is already in the defect band, allowing conduction in this subthreshold regime. A positive gate voltage further increases the current in these n-type transistors, while the small magnitude of $V_{th}$ indicates that this Fermi level is close to the defect band edge. We note that for all the different vacancy densities considered in our calculations, the defect band is centered closer to the conduction band (Table S1), consistent with the n-type transfer characteristics in experiment. The variability observed in experiment arises from slightly different alignments in the Fermi level of electrodes and defect bands in different transistor samples, but all the results point to a regime of small $V_{th}$. Apart from that, the presence of large negative $V_{th}$ observed in the transistors fabricated on pristine $MoS_2$ flakes (labelled 'P' in Figure 1b) could be attributed to extraneous effects such as trapped donors at the $MoS_2/SiO_2$ interface.[18]

The mechanism that we discuss here on how the $V_{th}$ of a $MoS_2$ transistor changes with vacancy density is consistent with a recent study that varies the vacancy density in $MoS_2$ monolayer films by controlling the chemical vapour deposition (CVD) growth process.[35] Here, we show that the vacancy density in mechanically exfoliated $MoS_2$ can also be easily controlled to tune the $V_{th}$ using CMOS-compatible processes. Importantly, the room temperature hydrogen treatment proposed here can be used to fine-tune the $V_{th}$ of any finished $MoS_2$ transistor (including those made with CVD-grown $MoS_2$ films), without degradation in performance.

The photoluminescence (PL) intensity of a $MoS_2$ flake with uniform thickness ought to be significantly higher wherever there are sulfur vacancies in the basal plane of $MoS_2$ due to the suppression of non-radiative recombination of excitons at the sulfur vacancies.[36] Through a series of PL analyses, we have verified that the sulfur and hydrogen treatments reduce and increase the amount of sulfur vacancies in the basal plane of $MoS_2$, respectively. Figure 4a shows the optical image of a piece of freshly-exfoliated (pristine) $MoS_2$ flake while Figure 4b shows the respective PL intensity map for the region marked in Figure 4a. Subsequently, the sulfur treatment was performed on the sample and Figure 4c shows the corresponding PL intensity map. After that, the sample was subjected to hydrogen treatment and its corresponding PL intensity map is shown in Figure 4d. It should be noted that all the PL intensity maps in Figure 4 share the same color intensity bar and as such, the intensity level measured is a representative indicator of the defect density throughout the $MoS_2$ surface after undergoing different processes. As expected, the PL intensity of the $MoS_2$ flake is more uniform after the sulfur treatment compared to when it was as-exfoliated. As shown in Figure 4c, the PL intensity is almost uniform throughout the MoS$_2$ flake, although it was not uniform even when it was as-exfoliated (Figure 4b) and the positions with higher PL intensity are believed to be where the sulfur vacancies present in the surface of the as-exfoliated MoS$_2$ flake. These observations confirm that the sulfur treatment is able to reduce the amount of sulfur vacancies that exist in the as-exfoliated MoS$_2$ flake. On the other hand, the PL intensity throughout the MoS$_2$ flake is observed to be non-uniform after the hydrogen treatment, thus indicating the creation of a large number of non-uniformly-distributed sulfur vacancies in the basal plane of the MoS$_2$ flake. This agrees well with our hypothesis that the hydrogen treatment is able to create a substantial amount of sulfur vacancies in the MoS$_2$ surface. In addition, Figure 4e shows the PL spectrum of a typical MoS$_2$ flake which contains 2 peaks and one of them positioned at 1.83 eV coincides with the monolayer emission peak. For the MoS$_2$ flake, the PL peaks became much weaker after the sulfur treatment, but the intensity increases after undergoing the hydrogen treatment (Figure 4e). Similar results were observed for MoS$_2$ flakes with different number of layers (Figure S7). Furthermore, we also compare the PL intensity map of a piece of exfoliated MoS$_2$ flake before and after the hydrogen treatment (Figure S8). In short, the changes in PL intensity for the MoS$_2$ flake after each treatment can be explained by the changes in the amount of sulfur vacancies in the MoS$_2$ surface and the overall PL analysis results corroborate our hypothesis that the sulfur / hydrogen treatment repairs / creates sulfur vacancies in the basal plane of MoS$_2$. More importantly, the hypothesis is supported by an X-ray photoelectron spectroscopy (XPS) study which demonstrates minor stoichiometry changes in the surface of MoS$_2$ (see Supplementary Information S7 for details).

As mentioned earlier, our sulfur treatment involves annealing the MoS$_2$ sample at ~435 °C for 2 h which is expected to repair the sulfur vacancies in the MoS$_2$ through bonding of sulfur atoms to the dangling Mo bonds. It is important to note that the annealing process must be conducted in a sulfur-rich environment, and failure to do so would result in the introduction of more rather than fewer sulfur vacancies. In Figure 5, we demonstrate the impact of annealing the MoS$_2$ sample without sulfur vapor. A sample (exfoliated MoS$_2$ flakes on a p$^+$ Si / SiO$_2$ substrate) was first sulfur-treated. Its optical image and PL intensity map over the energy range of 1.7 to 2.0 eV are shown in Figure 5a. Subsequently, the sample was annealed using the same experimental setup and conditions as the sulfur treatment except that no sulfur powder was loaded and no Ar gas was introduced. In other words, the sample in this case was simply annealed at ~435 °C for 2 h in vacuum (2 x $10^{-4}$ mbar). The few-layer $MoS_2$ flake shows uniform PL intensity (Figure 5a) and weak PL signal (Position 1 in Figure 5c) after being sulfur-treated similar to the case shown in Figure 4. In contrast, the same few-layer $MoS_2$ flake shows very non-uniform PL intensity (Figure 5b) and much stronger PL signal (Positions 2&3 in Figure 5c) after being annealed in vacuum in the absence of sulfur vapor, which indicates that many sulfur vacancies have been generated by the vacuum annealing process. This observation correlates well with the findings of Nan *et al.* that a 1 hour vacuum anneal at 350 °C is sufficient to generate a large number of sulfur vacancies in $MoS_2$ and results in a 6-fold enhancement of PL intensity.[36] On the other hand, our proposed hydrogen treatment is capable of creating multiple sulfur vacancies in the $MoS_2$ surface although not as many as the vacuum annealing process as evidenced by the PL and XPS studies. It is worth noting that the hydrogen treatment is neither a high-temperature nor a high-pressure process but simply a process conducted at room temperature and at atmospheric pressure.

**Experimental**

**Fabrication of back-gated $MoS_2$ field-effect transistors.** $MoS_2$ flakes were first exfoliated from molybdenite crystal (SPI supplies®) on an oxidized degenerately p-doped silicon substrate with 285 nm thick $SiO_2$. For this study, only $MoS_2$ flakes with thickness of ~5 nm and uniform width (2-3 μm) were selected. All selected $MoS_2$ flakes are either pristine, sulfur-treated, or hydrogen-treated. Subsequently, each sample (selected $MoS_2$ flakes on a $p^+$ Si / $SiO_2$ substrate) was then spin-coated with a 200 nm thick layer of polymethylmethacrylate (PMMA) 950 A4 (Microchem Inc.) and baked at 120 °C in an oven for 15 min. Subsequently, the source/drain contacts were delineated using electron beam lithography and metallized with Ti/Au (5/45 nm). It was followed by a 12-h lift-off process in acetone. For all $MoS_2$ devices in this work, the dimensions were kept constant, where the channel length and contact length are 1 μm and the channel width and contact width are about 2-3 μm, being the natural width of the exfoliated

MoS$_2$ flakes. Electrical measurements for all MoS$_2$ transistors were conducted at room temperature in a high vacuum chamber (10$^{-6}$ mbar) to avoid unnecessary interaction with moisture in ambient.[37]

**Sulfur treatment of MoS$_2$ flakes.** The sample (exfoliated MoS$_2$ flakes on a p$^+$ Si / SiO$_2$ substrate) was first loaded into a test tube containing 500 milligrams of sulfur powder at the closed end. The distance between the sample and sulfur powder is ~6 cm. The test tube was then loaded into a tube furnace with the sulfur powder positioned at the center of heating zone (Figure S2 in Supplementary Information). After that, the tube furnace was evacuated to a base pressure of ~2 x 10$^{-4}$ mbar. Subsequently, the furnace temperature was ramped up to the melting point of sulfur (445 °C) at a total pressure of 3 x 10$^{-1}$ mbar and held for 2 h. Throughout the annealing process, Argon gas flow (16 sccm) was introduced to control the diffusion rate of sulfur vapor and the sample temperature was measured to be ~435 °C.

**Hydrogen treatment of MoS$_2$ flakes.** The sample (exfoliated MoS$_2$ flakes on a p$^+$ Si / SiO$_2$ substrate) was first loaded into a vacuum chamber. The chamber was then evacuated to a pressure of 1 x 10$^{-6}$ mbar. Subsequently, the chamber was filled with pure hydrogen gas to one atmospheric pressure (1 x 10$^3$ mbar) forming a hydrogen-rich environment. The chamber conditions (1 atm, 25 °C) was retained for 2 h before unloading the sample.

**Density functional theory (DFT) calculations.** First-principles calculations were performed using DFT calculations with the local density approximation (LDA) for the exchange-correlation functional as implemented in the plane-wave code VASP.[38] The standard scalar relativistic projector-augmented wave (PAW) potentials were employed throughout the calculation. A plane wave kinetic energy cut-off of 400 eV is used (similar band structures are obtained with an increased cut-off of 500 eV) and a vacuum thickness of 16 Å is included between slabs. The Gamma-centered k-point mesh of 15×15×1 is used in the perfect bilayer MoS$_2$ and Ti-covered-MoS$_2$ unit cell self-consistent-field calculations. The k-point meshes of 3×3×1 and 5×5×1 are used in the 5×5×1 and 3×3×1 supercell MoS$_2$ for different S vacancy density simulations, respectively. The results thus obtained are converged relative to higher energy cut-offs and denser k-point meshes (see Supplementary Information S5 for details). All of the atomic coordinates and in-plane lattice constants are optimized with the conjugate gradient algorithm. The structures are considered as relaxed when the maximum component of the Hellmann−Feynman force acting on each ion is less than 0.01 eV/Å. For the Ti-covered-MoS$_2$ system, 4 layers of Ti atoms are used and the lattice constants are set to that of the optimized bilayer MoS$_2$ system. In the self-consistent calculation, the convergence threshold for energy is set to $10^{-5}$ eV.

**PL analysis.** For this study, all PL measurements were conducted using a WITec alpha 300R confocal Raman system with a 532 nm laser excitation source and a laser spot size of ~320 nm was used. The laser power at the sample was kept below 0.1 mW to avoid laser-induced damage at the sample.[39] For PL mappings, the sample was placed on a piezostage and scanned with a step size of 100 nm, unless otherwise specified.

**Conclusion**

We have introduced two facile approaches, namely sulfur treatment and hydrogen treatment, which allow one to fine-tune the threshold voltage of MoS$_2$ FETs while being Si-CMOS process compatible at the same time. Using a sulfur treatment process to reduce the number of sulfur vacancies on as-exfoliated MoS$_2$, we demonstrate that the $V_{th}$ of fabricated FETs can be shifted into the positive regime. On the other hand, left shifting of the $V_{th}$ by populating the exposed MoS$_2$ channel with sulfur vacancies can be achieved by a hydrogen treatment. In addition, we show that not only could $V_{th}$ be tuned in both directions, the treatment is also able to marginally improve the intrinsic electrical properties of the MoS$_2$, with the treated FET exhibiting respectable electrical performance - high on-off current ratio ($> 10^5$) and 2-fold enhancement in electron mobility; the performance does not degrade even after multiple treatments. Hence as such, our work provides a means to adjust the critical threshold voltage parameter needed in the development of MoS$_2$ as a channel material for next generation transistors.

**Acknowledgements**

This project is supported by grant R-263-000-A76-750 from the Faculty of Engineering, NUS and grants NRF2011NRF-CRP002-050 and NRF-NRFF2013-07 from the National Research Foundation, Singapore. Calculations were performed on the NUS Graphene Research Centre Computational Cluster. We acknowledge NRF for funding through the Mid-Size Centre grant at CA2DM.

FIGURES

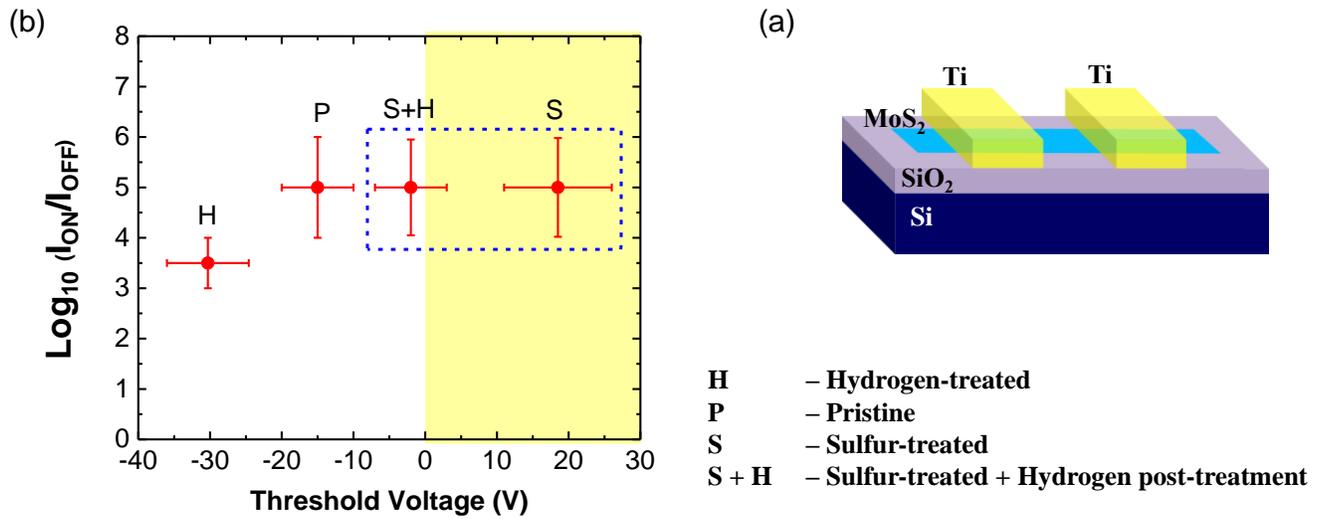

**Figure 1.** (a) Schematic showing the structure of a back-gated $MoS_2$ field-effect transistor used throughout this work. (b) The on-off current ratio versus the extracted threshold voltage of all $MoS_2$ field-effect transistors with the legend indicating different treatment processes. The average threshold voltage for the same group of transistors that were first fabricated on sulfur-treated $MoS_2$ flakes is left-shifted from 18.5 ± 7.5 V to -2 ± 5 V following the hydrogen post-treatment without any degradation being observed in the on/off current ratio as indicated by the blue dotted box, in other words, the $V_{th}$ of a $MoS_2$ transistor is adjustable by controlling the amount of sulfur vacancies in the uncovered $MoS_2$ channel.

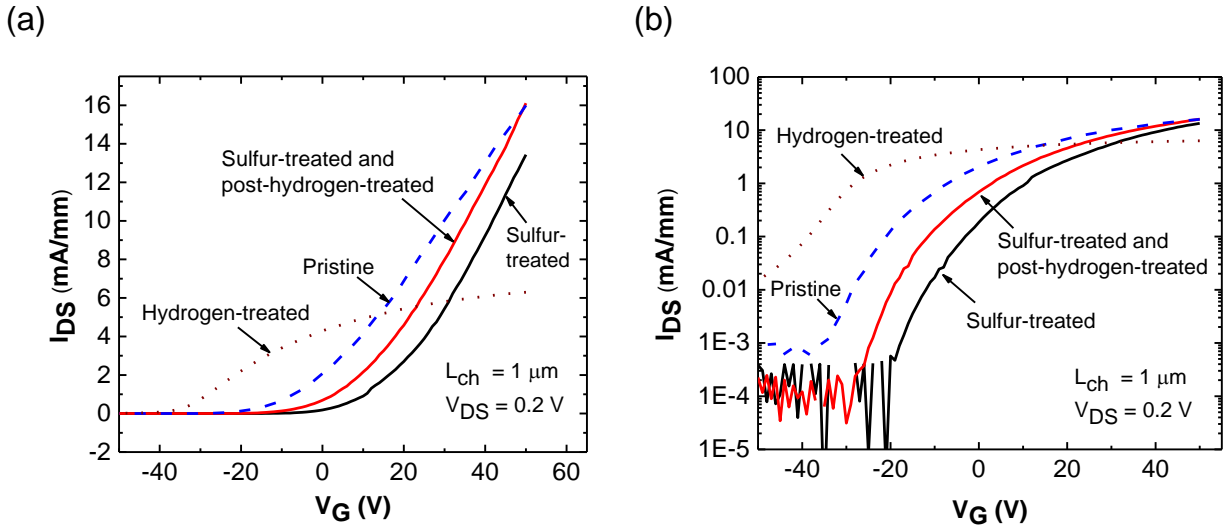

**Figure 2.** $I_D$-$V_G$ characteristics of a typical back-gated $MoS_2$ transistor that was first fabricated on a sulfur-treated $MoS_2$ flake and followed by a hydrogen-treatment in both linear (a) and logarithmic scale (b). The extracted threshold voltage of the as-fabricated transistor on the sulfur-treated $MoS_2$ flake is 12 V and shifts to 0 V after the hydrogen-treatment with no significant changes observed in terms of on/off current ratio and field-effect mobility (difference ~ 3%). For comparison purposes, $I_D$-$V_G$ characteristics of 2 typical back-gated $MoS_2$ transistors fabricated on either pristine or hydrogen-treated $MoS_2$ flake of similar thickness are included.

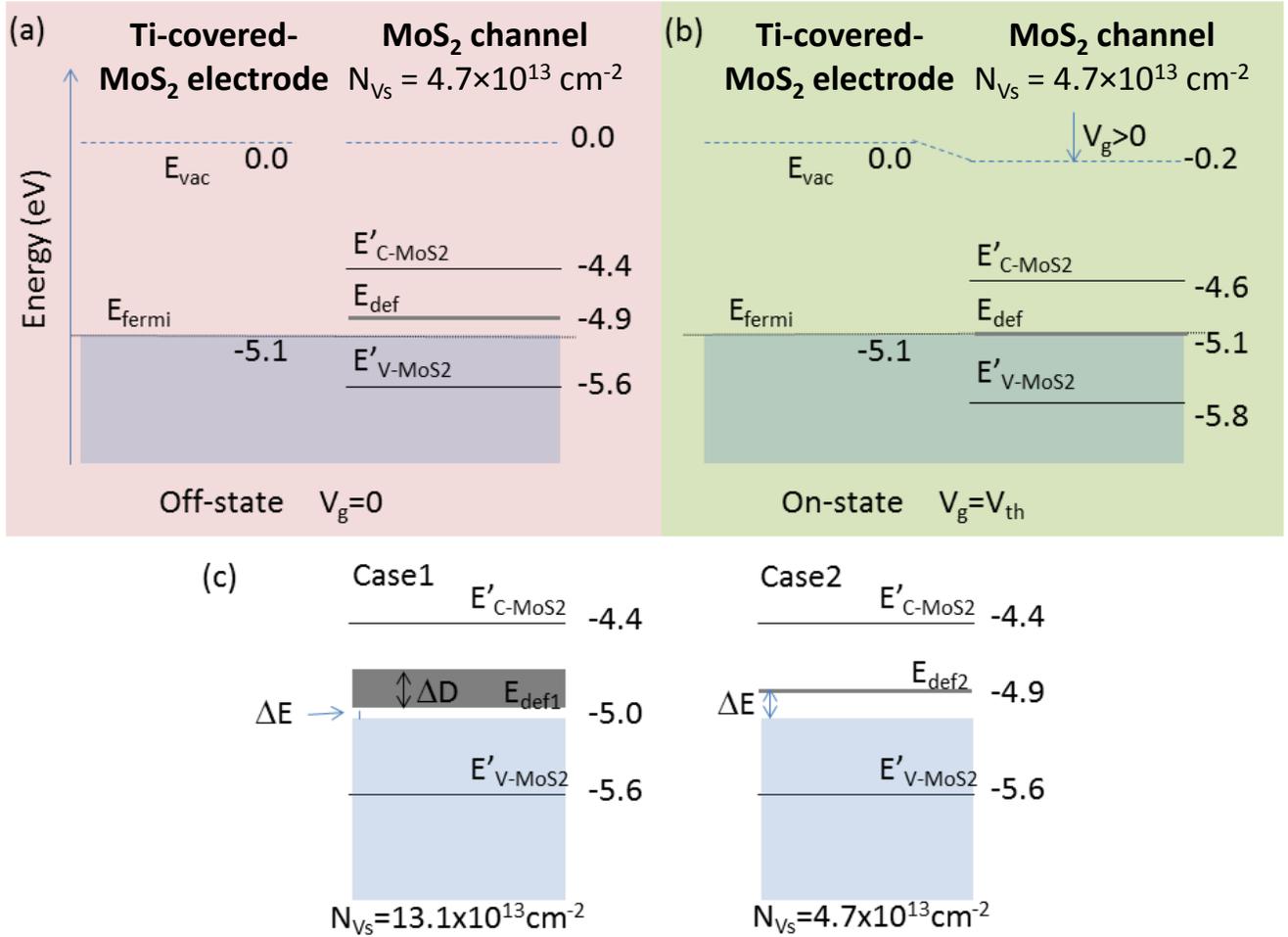

**Figure 3.** (a) Energy band diagram of the MoS$_2$ device in the off-state ($V_g = 0$) based on DFT calculations, $E_{vac}$ and $E_{fermi}$ respectively represent the vacuum level and the Fermi energy in Ti-covered-MoS$_2$ electrode. $E'_{C\text{-}MoS2}$, $E_{def}$ and $E'_{V\text{-}MoS2}$ represent the conduction band minimum, bottom edge of defect band and valence band maximum of MoS$_2$ channel with a density of sulfur vacancy ($N_{Vs}$) of $4.7\times10^{13}$ cm$^{-2}$, respectively. The energies are referenced to the vacuum level, which is set to zero. (b) Energy band diagram of the MoS$_2$ device in the on-state ($V_g = V_{th}$); the back gate voltage can shift the energy band of the MoS$_2$ channel, while that of Ti-covered-MoS$_2$ electrode is fixed. (c) Energy band diagram of MoS$_2$ channel for different $N_{Vs}$: Case1 and Case2 represent the systems where the $N_{Vs}$ are $13.1\times10^{13}$ and $4.7\times10^{13}$ cm$^{-2}$, respectively. The value of defect band width $\Delta D$ for different $N_{Vs}$ is listed in Table S1. We note that for negative $V_{th}$, the device is already in the on-state at $V_g = 0$ (see red curve in Figure 2a).

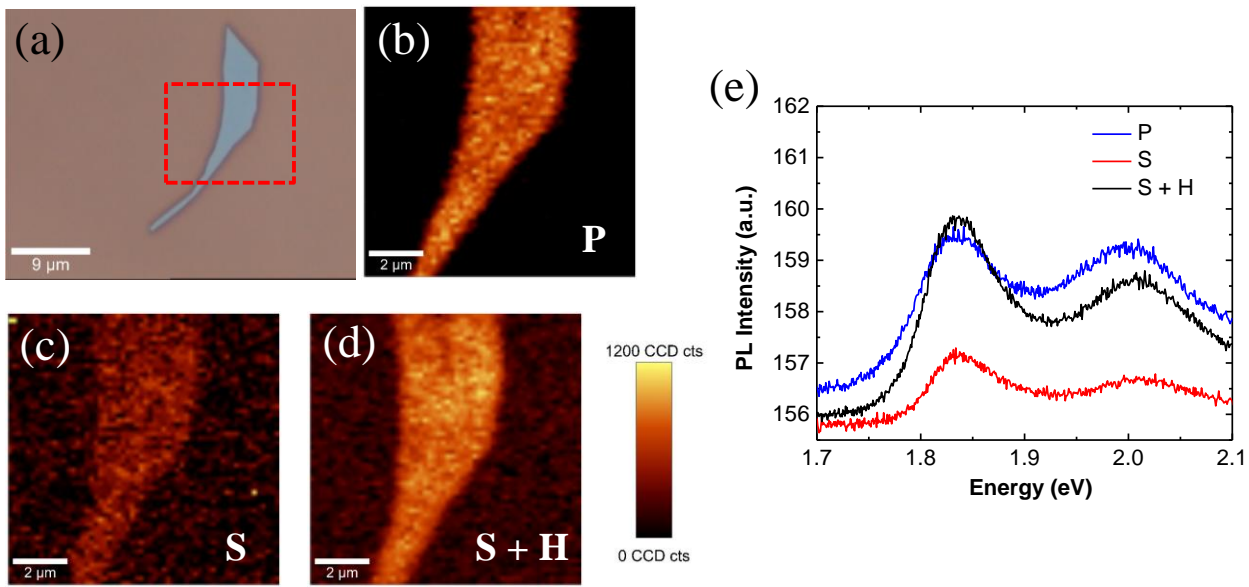

**Figure 4.** Verification of reduction and creation of sulfur vacancies in the basal plane of $MoS_2$ by sulfur and hydrogen treatments, respectively, through Photoluminescence (PL) analysis. (a) Optical image of a uniform few-layer $MoS_2$ flake exfoliated on a $p^+$ Si / $SiO_2$ substrate. (b-d) PL intensity maps of the typical $MoS_2$ flake within the marked region in (a) when it was as-exfoliated (P), sulfur-treated (S), and following the hydrogen treatment (S + H). All PL maps share the same color intensity bar. (e) PL spectra of the $MoS_2$ flake taken at different stages as indicated.

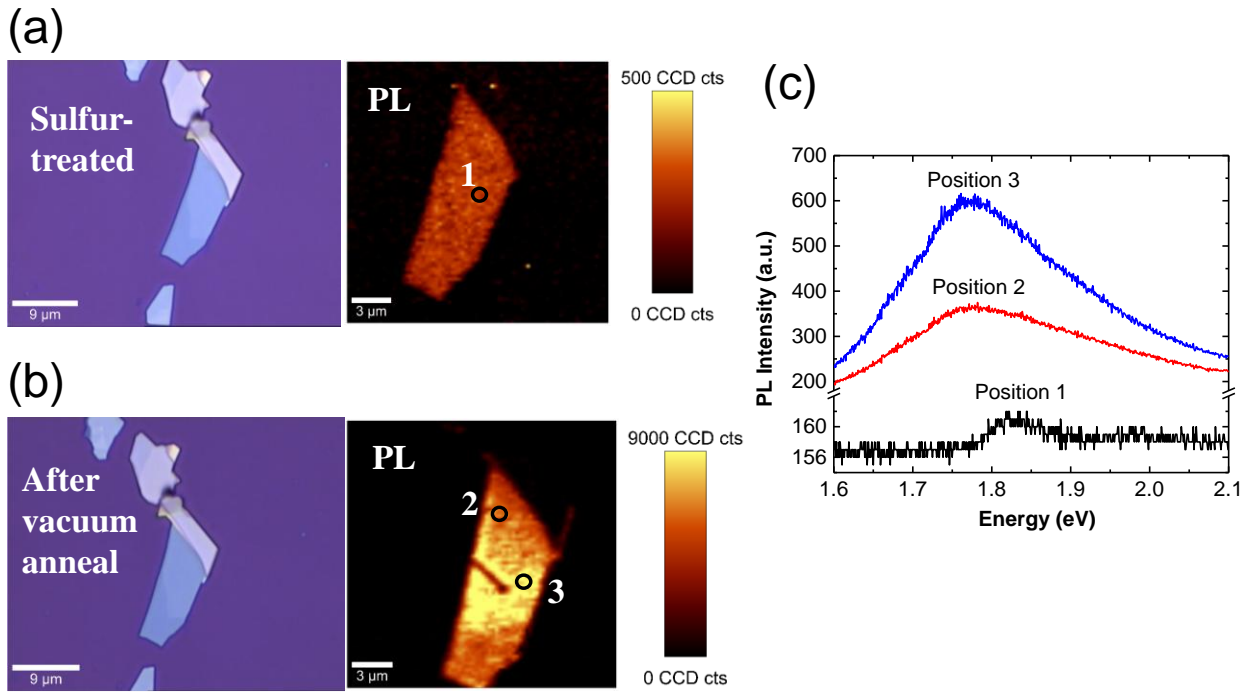

**Figure 5.** Impact of annealing the MoS$_2$ sample without sulfur vapor. (a) Optical image of a sulfur-treated MoS$_2$ flake and its corresponding PL intensity map. (b) Optical image of the same MoS$_2$ flake and its corresponding PL intensity map after being annealed in the absence of sulfur vapor. (c) PL spectra of the MoS$_2$ flake taken at different positions as indicated.

# Supplementary Information

**S1. Thickness confirmation for MoS$_2$ flakes used in this study**

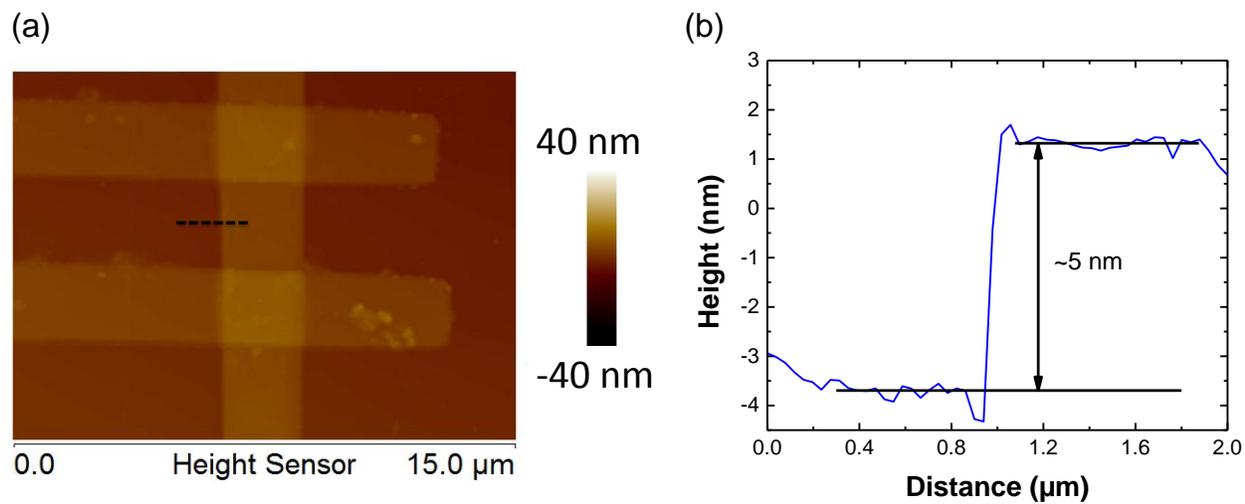

**Figure S1.** (a) Typical atomic force microscopy (AFM) topography image of an exfoliated MoS$_2$ strip. (b) Height profile along the dotted line in (a).

## S2. Experimental setup of sulfur treatment for $MoS_2$ flakes

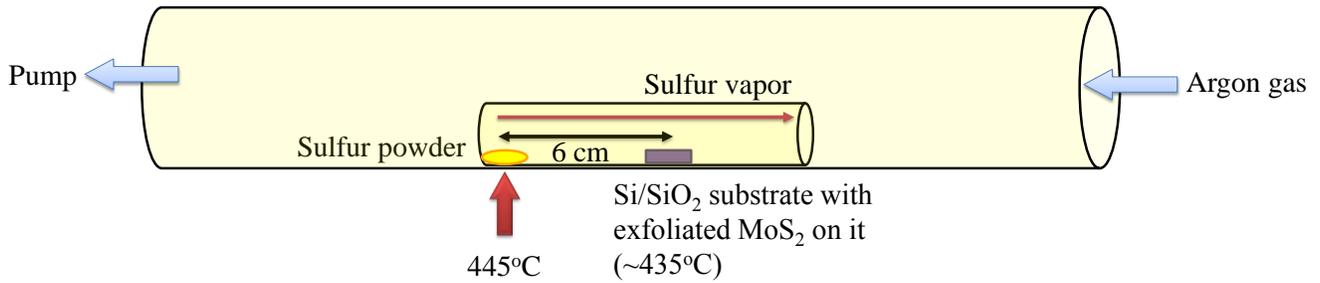

**Figure S2.** Schematic showing the experimental setup for sulfur treatment of $MoS_2$ flakes. The sample (exfoliated $MoS_2$ flakes on a $p^+$ Si / $SiO_2$ substrate) was placed into a test tube and 6 cm away from the sulfur powder (500 milligrams) at the closed end. The test tube was positioned in the tube furnace such that sulfur powder is at the center of heating zone. The furnace temperature was ramped up to the melting point of sulfur (445 °C) at a total pressure of $3 \times 10^{-1}$ mbar and held for 2 h. Throughout the annealing process, Argon gas flow (16 sccm) was introduced to control the diffusion rate of sulfur vapor and the sample temperature was ~435 °C.

## S3. Transfer characteristics of transistors fabricated on different types of MoS$_2$

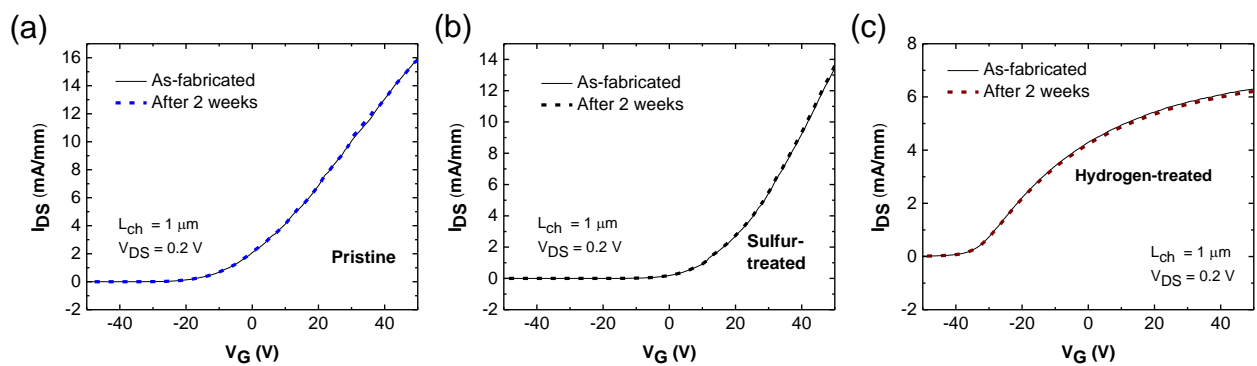

**Figure S3.** $I_D$-$V_G$ characteristics of 3 typical back-gated MoS$_2$ transistors fabricated on (a) pristine, (b) sulfur-treated, and (c) hydrogen-treated MoS$_2$ flake, during the day when they were freshly-fabricated and after 2 weeks of storage in ambient conditions (25 °C, 1 atm).

**S4. Output characteristics of transistors fabricated on different types of MoS$_2$**

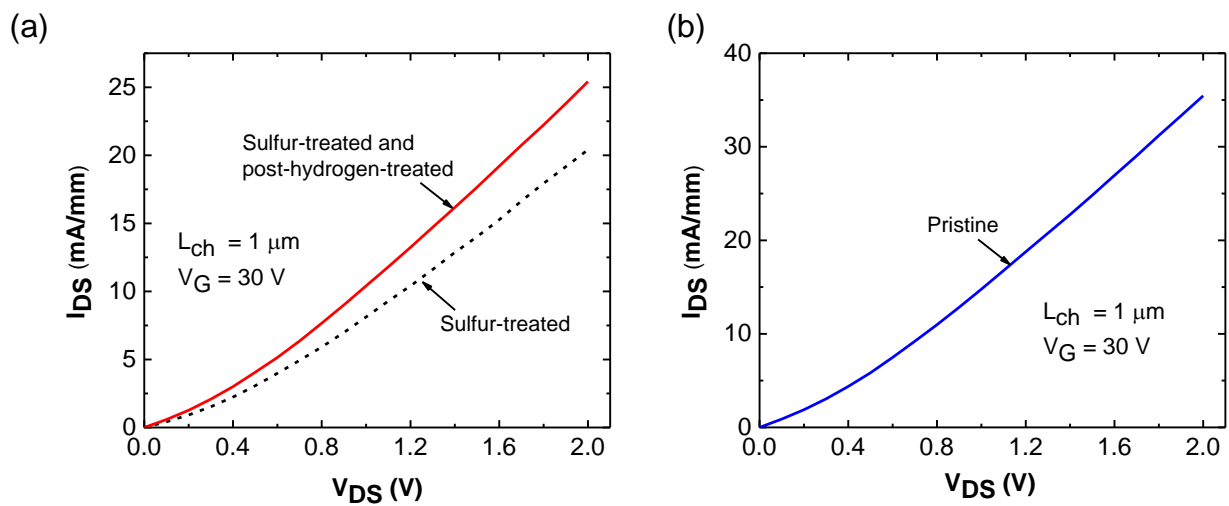

**Figure S4.** (a) $I_D$-$V_D$ characteristics of a typical back-gated MoS$_2$ transistor that was first fabricated on a sulfur-treated MoS$_2$ flake and followed by a hydrogen-treatment. (b) $I_D$-$V_D$ characteristics of a typical back-gated MoS$_2$ transistor that was fabricated on a pristine MoS$_2$ flake.

## S5. Band structure of Ti-covered-MoS$_2$ electrode and bilayer MoS$_2$

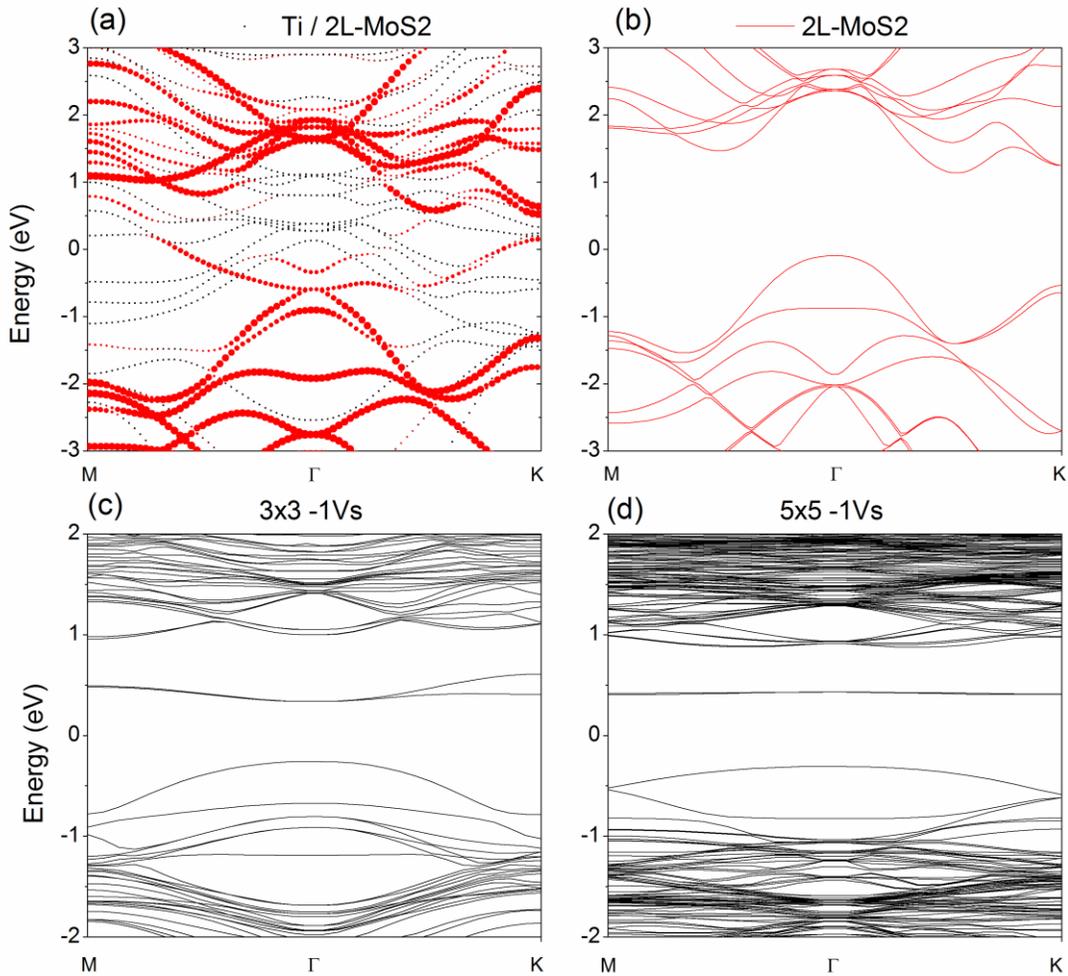

**Figure S5.** Electronic band structure of (a) Ti-covered-perfect-bilayer-MoS$_2$ unit cell, (b) perfect bilayer MoS$_2$ unit cell, (c) imperfect bilayer MoS$_2$ with one sulfur vacancy in 3×3×1 supercell, and (d) imperfect bilayer MoS$_2$ with one sulfur vacancy in 5×5×1 supercell. The characteristic band of MoS$_2$ is shown in red dots for the Ti-covered-MoS$_2$ combined system. We assigned the conduction band minimum and valence band maximum of the Ti-covered-perfect-bilayer-MoS$_2$ shown in (a) according to the band shape of the isolated MoS$_2$ shown in (b), and neglect the influence of the hybrid state (near Fermi level) induced by the adjacent metal.

**Table S1**: Energy eigenvalues (in eV) for the Fermi level ($E_{fermi}$), valence band maximum ($E_{VBM}$), conduction band minimum ($E_{CBM}$), bottom edge of defect band ($E_{def}$) and defect band width ($\Delta D$) for the perfect and defect containing systems. 3×3 -1Vs refers to one sulfur vacancy in the 3×3 bilayer $MoS_2$ supercell. All the energies are referenced to the vacuum level (which is set to 0 eV). $\Delta E$ is the energy shift required to align the $E_{def}$ in $MoS_2$ channel (bare $MoS_2$) to the Fermi level in the Ti-covered-$MoS_2$ electrode (-5.1 eV).

| Bilayer MoS₂ Supercell | 2×2 -1Vs | 3×3 -1Vs | 4×4 -1Vs | 5×5 -1Vs | 6×6 -1Vs | Perfect |
|---|---|---|---|---|---|---|
| Density of sulfur vacancy (×10¹³ cm⁻²) | 29.6 | 13.1 | 7.4 | 4.7 | 3.2 | 0 |
| $E_{fermi}$ (eV) | -5.43 | -5.34 | -5.33 | -5.31 | -5.31 | -5.06 |
| $E_{VBM}$ (eV) | -5.54 | -5.60 | -5.62 | -5.62 | -5.63 | -5.62 |
| $E_{CBM}$ (eV) | -4.36 | -4.39 | -4.45 | -4.44 | -4.45 | -4.46 |
| Defect band edge, $E_{def}$ (eV) | -5.23 | -5.00 | -4.95 | -4.91 | -4.90 | -- |
| Center of defect band (eV) | -4.84 | -4.87 | -4.91 | -4.90 | -4.90 | -- |
| Defect band width, $\Delta D$ (eV) | 0.78 | 0.27 | 0.09 | 0.03 | 0.01 | -- |
| $\Delta E = E_{def} - (-5.1\ eV)$ (eV) | **-0.13** | **0.10** | **0.15** | **0.19** | **0.20** | -- |

**Table S2:** Energy values of the Fermi level ($E_{fermi}$), valence band maximum ($E_{VBM}$), conduction band minimum ($E_{CBM}$), bottom edge of defect band ($E_{def}$) and defect band width ($\Delta D$) for the 3×3 -1Vs and 6×6 -1Vs in different convergence parameters. All the energies are referenced to the vacuum level (which is set to 0 eV). $E_{cut}$ is the cutoff energy used in the calculation. Both set of values are almost the same, indicating the calculation is converged, and the trend of the defect band is robust with more converged parameters.

| Bilayer MoS₂ Supercell | 3×3 -1Vs | | 6×6 -1Vs | |
|---|---|---|---|---|
| Convergence parameters | $E_{cut}$: 400eV, k-point: 5×5×1 | $E_{cut}$: 500eV, k-point: 8×8×1 | $E_{cut}$: 400eV, k-point: 2×2×1 | $E_{cut}$: 500eV, k-point: 4×4×1 |
| $E_{fermi}$ (eV) | -5.341 | -5.342 | -5.310 | -5.309 |
| $E_{VBM}$ (eV) | -5.600 | -5.601 | -5.629 | -5.629 |
| $E_{CBM}$ (eV) | -4.386 | -4.388 | -4.454 | -4.454 |
| $E_{def}$ (eV) | -5.004 | -5.006 | -4.906 | -4.905 |
| $\Delta D$ (eV) | 0.272 | 0.272 | 0.011 | 0.011 |

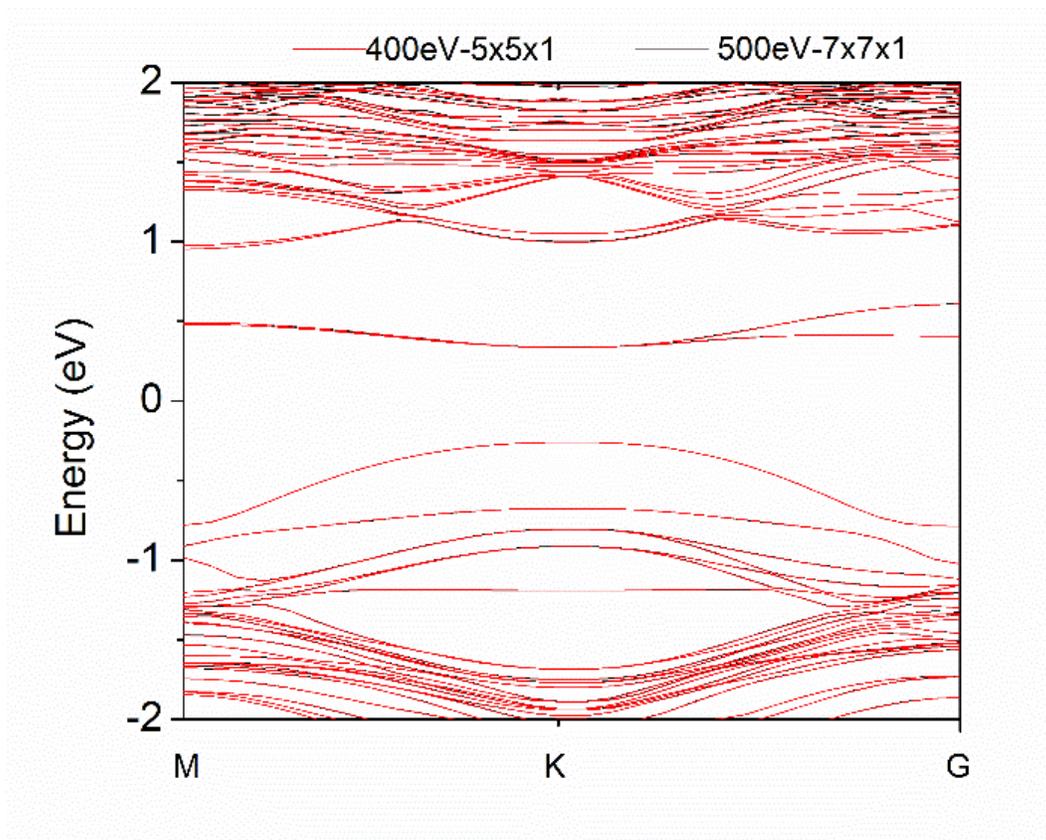

**Figure S6.** Electronic band structure of imperfect bilayer MoS$_2$ with one sulfur vacancy in 3×3×1 supercell, the red line and black line represent the convergent parameters used in this paper and a higher convergent parameters (500 eV energy cutoff for plane wave basic set and 7×7×1 Monkhorst-Pack *k*-point sampling), respectively. Both lines are almost overlapping, indicating the calculation is converged, and the trend of the defect band is robust with more converged parameters.

## S6. Photoluminescence (PL) study of surface-treated MoS$_2$ flakes

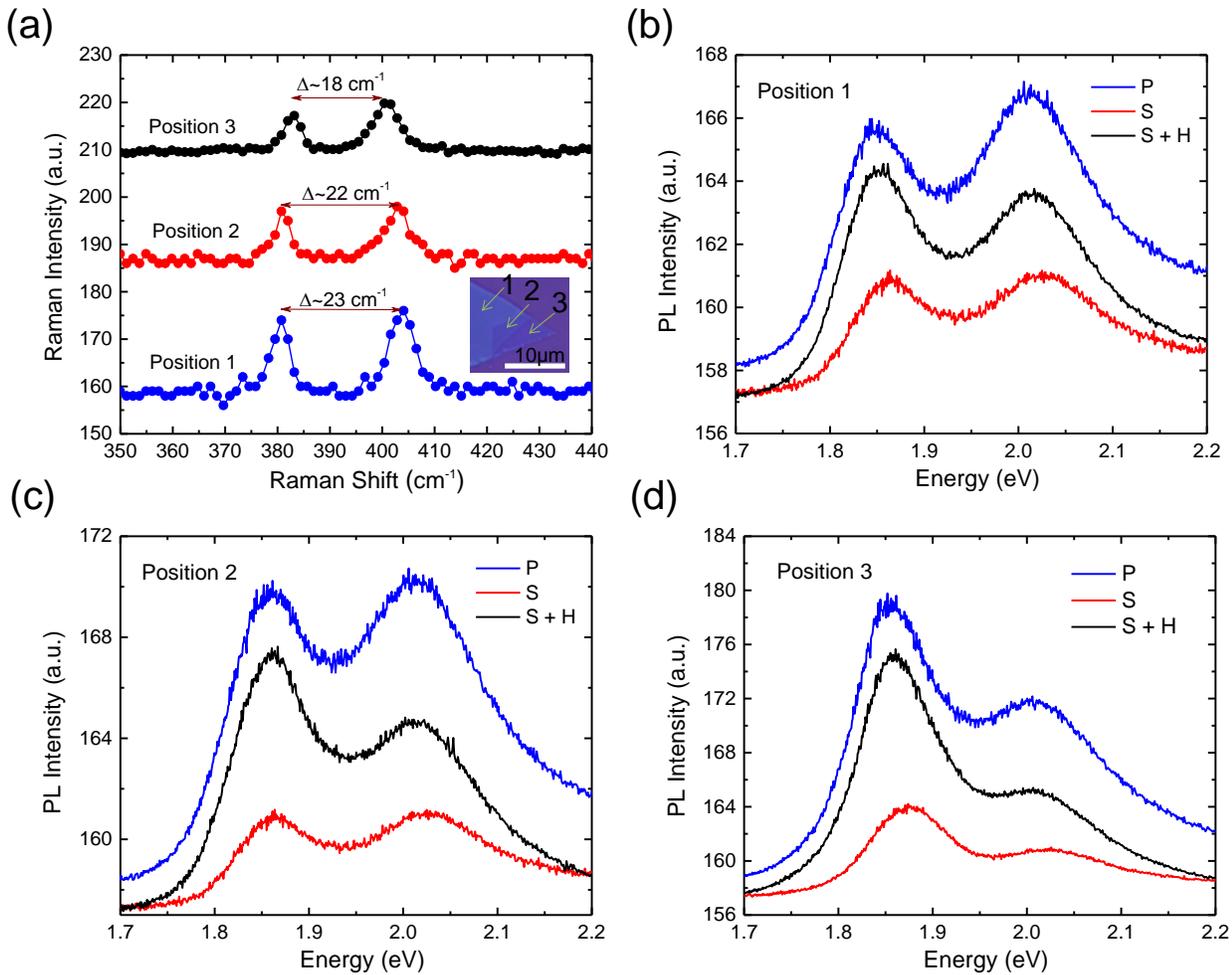

**Figure S7.** (a) Raman spectra of pristine MoS$_2$ flakes taken at different positions as indicated. Inset: Optical image of the MoS$_2$ flakes exfoliated on a p$^+$ Si / SiO$_2$ substrate. In particular, positions 2&3 represent few-layer MoS$_2$ while position 1 represents a monolayer MoS$_2$. We note that the Raman spectra of these MoS$_2$ flakes are the same in terms of both relative intensity and peaks' position when they were was pristine (P), sulfur-treated (S), and followed by the hydrogen treatment (S + H). (b-d) PL spectra of the MoS$_2$ flakes taken at different positions and stages (P, S, S+H) as indicated.

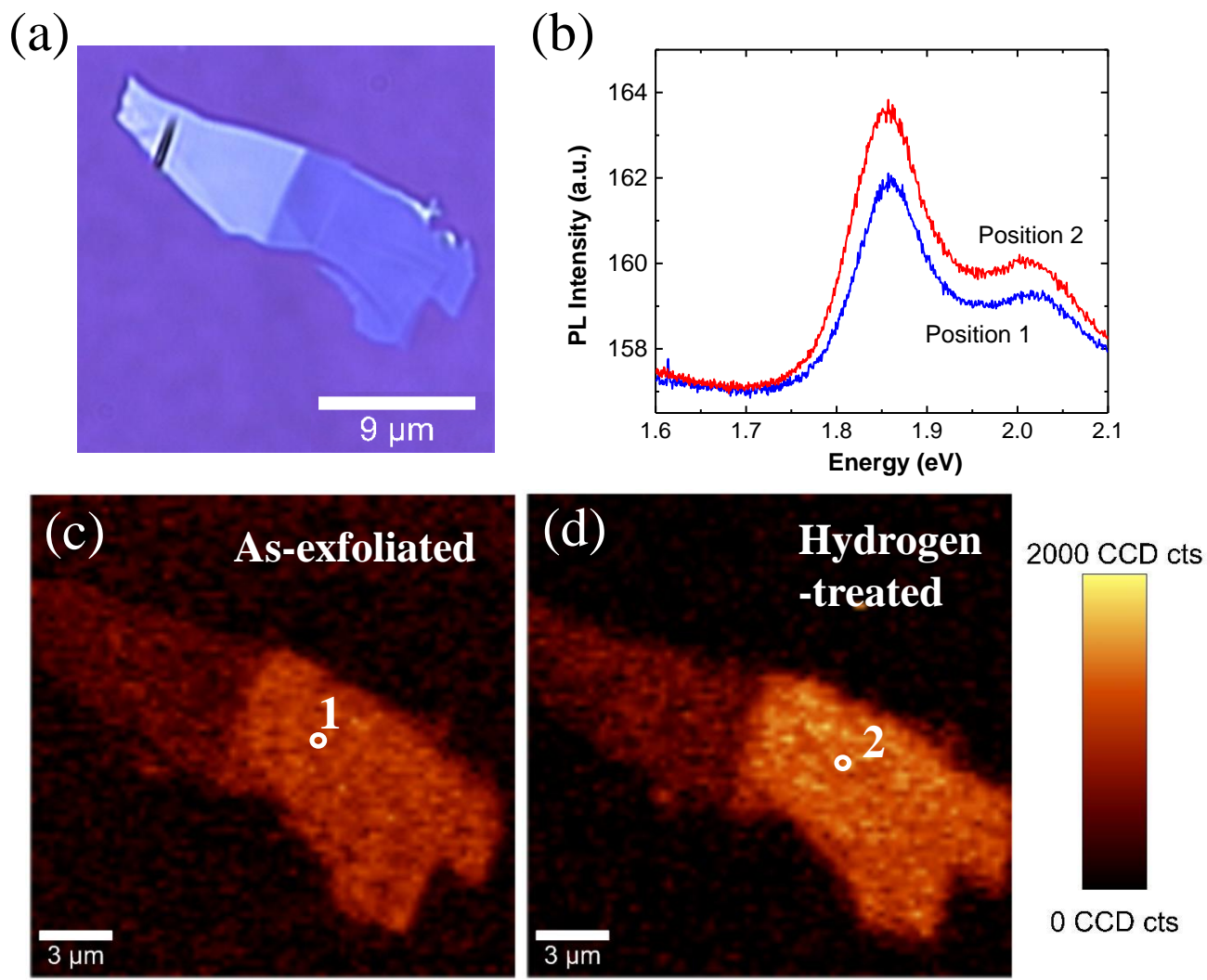

**Figure S8.** (a) Optical image of a MoS$_2$ flake exfoliated on a p$^+$ Si / SiO$_2$ substrate. (b) PL spectra of the MoS$_2$ flake taken at different positions as indicated in c & d. (c) The PL intensity map of the MoS$_2$ flake when it was as-exfoliated. (d) The PL intensity maps of the MoS$_2$ flake after it underwent the hydrogen treatment. All PL maps share the same color intensity bar.

## S7. Material composition study on the surface-treated MoS$_2$ sample

As mentioned in the main text, the proposed sulfur and hydrogen treatments are expected to repair and create sulfur vacancies in the basal plane of MoS$_2$, respectively. Here, we attest the hypothesis by investigating stoichiometry changes of the MoS$_2$ prior to and following the proposed treatments through an X-ray photoelectron spectroscopy (XPS) study. A piece of bulk MoS$_2$ flake was first cleaved from molybdenite crystal (SPI supplies®) and its XPS spectrum is shown in Figure S9 (labelled as P). It should be noted that all XPS spectra in this study were obtained by exciting the MoS$_2$ sample with an Mg Kα source (XR 50, SPECS GmbH), and detected with a PHOIBOS 150 Hemispherical Energy Analyzer equipped with a Delay Line Detector (SPECS GmbH) at a pass energy of 30 eV. Subsequently, the MoS$_2$ sample was subjected to the sulfur treatment and its XPS spectrum is plotted in Figure S9 too (labelled as S). The same MoS$_2$ sample was then exposed to the hydrogen treatment and its XPS spectrum is also included in Figure S9 (labelled as S+H) for comparison purposes. As can be seen in Figure S9, the Mo 3$d$ and S 2$p$ XPS spectra of the MoS$_2$ sample show slight changes after the sulfur treatment followed by the hydrogen treatment. Table S3 summarizes the stoichiometry (atomic ratio of Mo:S) of the same MoS$_2$ sample when it was pristine (P), sulfur-treated (S), and following the hydrogen treatment (S + H), which were identified through components' peak fitting (an example is shown in Figure S10). As expected, for the same MoS$_2$ sample, the stoichiometry changes from 1.89 to 1.96 after the sulfur treatment, which indicates a reduction of sulfur vacancies in the MoS$_2$ surface, and more importantly, stoichiometry of the same MoS$_2$ sample changes to 1.90 after the hydrogen treatment, which signifies an increase in sulfur vacancies in the MoS$_2$ surface. Overall, the results agree well with our hypothesis that the proposed sulfur / hydrogen treatment reduces / increases the amount of sulfur vacancies in the basal plane of MoS$_2$.

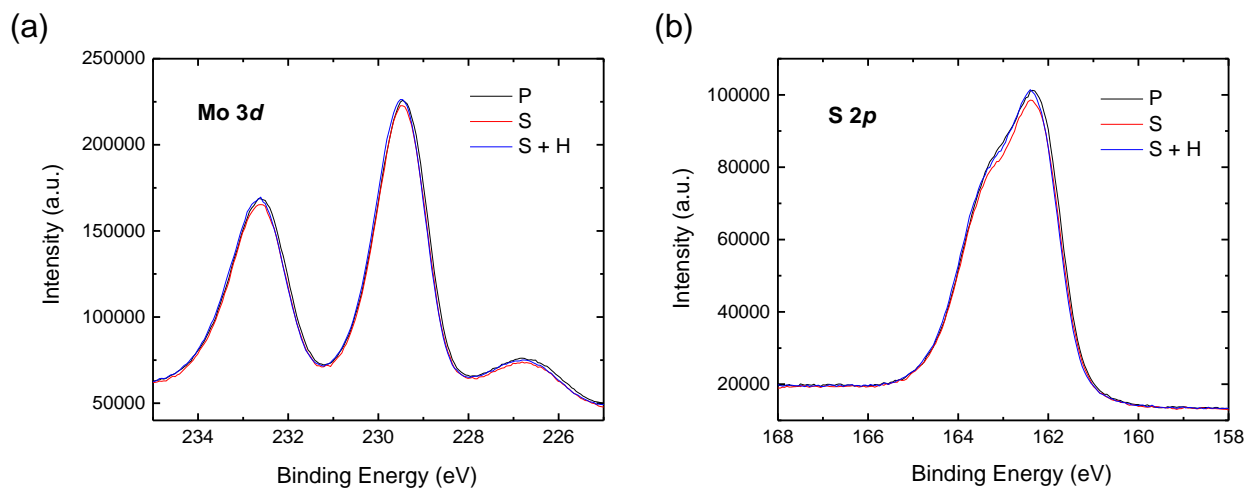

**Figure S9.** XPS spectra of the same MoS$_2$ sample when it was pristine (P), sulfur-treated (S), and followed by the hydrogen treatment (S + H). (a) Mo 3$d$ and (b) S 2$p$ XPS spectra of the MoS$_2$ sample.

**Table S3.** Atomic ratio of Mo:S of the same MoS$_2$ sample when it was pristine (P), sulfur-treated (S), and following the hydrogen treatment (S + H).

| Sample | Atomic ratio of Mo:S |
|--------|----------------------|
| P      | 1 : 1.89             |
| S      | 1 : 1.96             |
| S + H  | 1 : 1.90             |

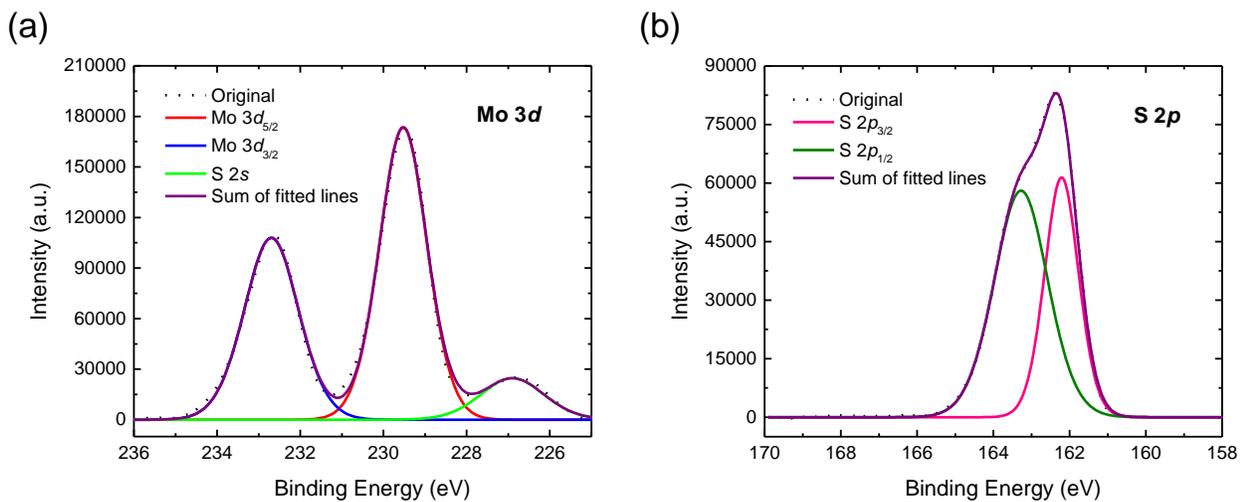

**Figure S10.** (a) Mo 3*d* and (b) S 2*p* XPS spectra of the "S + H" MoS$_2$ sample with curve fits to spectral components attributed to molybdenum (Mo 3$d_{3/2}$, Mo 3$d_{5/2}$) and sulfur (S 2*s*, S 2$p_{1/2}$ and S 2$p_{3/2}$) after Shirley background subtraction.